# Biased Power Spectrum and Bispectrum for an Ensemble of Three-Dimensional Scale Free Numerical Simulations

J. N. Fry

*Department of Physics, University of Florida, Gainesville FL 32611*

and Adrian L. Melott and Sergei F. Shandarin

*Department of Physics and Astronomy,*

*University of Kansas, Lawrence, KS 66045*

## ABSTRACT

We examine the effect of a threshold bias on the power spectrum and the bispectrum in an ensemble of numerical simulations (Gaussian initial perturbations with power law spectra $P(k) \sim k^n$, $n = +1, 0, -1, -2$) and compare our results with theoretical predictions. Our simulations are evolved sufficiently that on the scale where we apply the threshold the rms fluctuation has developed significantly into the nonlinear regime. Thus, predictions based on perturbation theory do not necessarily apply. Nevertheless, we find our results for the power spectrum, biased power simply amplified by a numerical factor, follow predicted trends, far beyond the regime where perturbation theory is expected to be valid.

We find that the biased bispectrum continues to follow the so-called hierarchical form, with reduced three-point amplitude $Q \approx 1$ in the strongly nonlinear regime, independent of initial spectrum. In the quasi-linear perturbative regime the three-point amplitude depends on configuration shape, a behavior that is found to give useful information about the amount of bias without information about the unbiased matter distribution.

*Subject headings:* methods: numerical

## 1. Introduction

Soon after Kaiser (1984) showed that the statistical properties of Abell clusters could be at least qualitatively understood as properties of high density peaks in the distribution of galaxies, it was recognized that the galaxy distribution itself might be similarly biased relative to the underlying distribution of mass. A bias in galaxy formation makes the statistical properties of the galaxy distribution different from those of the underlying mass distribution. In the intervening years a large number of statistical and physical bias mechanisms have been proposed, which usually but not always lead to a galaxy distribution more strongly clustered than the matter distribution. Results can be summarized by the bias parameter $b$, where the correlation function of the biased galaxy distribution $\xi_g$ is related to that of the matter distribution $\xi$ by $\xi_g = b^2 \xi$. It is almost axiomatic that some bias is necessary to reconcile the smoothness of the microwave background and the observed clustering strength of galaxies.

Theoretical calculations of what to expect from bias in galaxy formation are confined mainly to the perturbative regime. To study the further implications when fluctuation amplitudes evolve into the nonlinear regime the standard technique is numerical simulation. Melott & Shandarin (1993) have investigated the power spectrum and Fry, Melott, & Shandarin (1993) the bispectrum of the mass distribution at various evolution stages for an ensemble of three-dimensional numerical simulations for a series of scale-free initial conditions. In this paper we study the power spectrum and bispectrum after applying a threshold bias to the same ensemble of models (Gaussian initial perturbations with power law spectra $P(k) \sim k^n$, $n = +1, 0, -1, -2$). In Section 2 we review the behavior of the power spectrum and the bispectrum expected in perturbation theory, including effects of gravity and nonlinear bias. Section 3 presents our results from numerical simulations. Section 4 contains a discussion of what we have learned.

## 2. Theoretical Predictions
### 2.1 Gravitational Instability

The large-scale distribution of matter in the universe is shaped by the action of gravity on small initial seed fluctuations, usually taken to be Gaussian. Gravitational instability is nonlinear, thus precluding a completely analytic treatment. However, on sufficiently large scales, density fluctuations are weak enough that perturbation theory often suffices. To linear order, the fluctuation amplitude $\tilde{\delta}(\boldsymbol{k}, t)$, the Fourier transform of the density contrast $\delta = [\rho(\boldsymbol{x}, t) - \bar{\rho}]/\bar{\rho}$, grows by an overall scale factor, $\tilde{\delta} = A(t)\tilde{\delta}_0(\boldsymbol{k})$, where $\tilde{\delta}_0(\boldsymbol{k})$ is the amplitude at some early time $t_0$ (see Peebles 1980, § 10–13). For the canonical model, matter dominated, $\Omega = 1$, $\Lambda = k = p = R^{-2} = 0$, this factor is $A(t) \sim t^{2/3} \sim a(t)$, where $a(t)$ is the cosmological expansion factor, and normalization can be chosen so that $A(t) = a(t)$. Thus, perturbation theory predicts to leading order the power spectrum $P(k) = \langle |\tilde{\delta}(\boldsymbol{k})|^2 \rangle$ simply grows with time,

$$P(k) = a^2(t) P_0(k). \tag{1}$$

For Gaussian initial conditions, all higher order irreducible correlations for $\tilde{\delta}$ vanish initially. However, in a nonlinear theory, even for a Gaussian initial distribution, nonlinearities induce nonvanishing higher order correlations for all orders (see Fry 1984, Goroff $et$ $al.$ 1986, Bernadeau 1992). For the three-point function in particular, $B_{123} = \langle \tilde{\delta}(\boldsymbol{k}_1)\tilde{\delta}(\boldsymbol{k}_2)\tilde{\delta}(\boldsymbol{k}_3) \rangle$ for $\sum \boldsymbol{k}_i = 0$, gravitational instability gives

$$B_{123} = \mathcal{B}_{12} + \mathcal{B}_{13} + \mathcal{B}_{23} \tag{2}$$

where

$$\mathcal{B}_{ij} = \left[ \frac{10}{7} + \hat{\boldsymbol{k}}_i \cdot \hat{\boldsymbol{k}}_j \left( \frac{k_i}{k_j} + \frac{k_j}{k_i} \right) + \frac{4}{7} \left( \hat{\boldsymbol{k}}_i \cdot \hat{\boldsymbol{k}}_j \right)^2 \right] P(k_i) P(k_j). \qquad (3)$$

Note that since $P \sim a^2(t)$, $B \sim a^4$.

We can remove the main dependence on scale, on time, and on the initial power spectrum by using the hierarchical amplitude $Q$, defined as

$$Q_{123} = \frac{B_{123}}{P_1 P_2 + P_1 P_3 + P_2 P_3}. \qquad (4)$$

In a pure hierarchical model, $Q$ would be strictly constant. For equilateral triangle configurations $k_i/k_j = 1$ and $\hat{\boldsymbol{k}}_i \cdot \hat{\boldsymbol{k}}_j = -1/2$ for all pairs, and equation (2) gives $Q_{123} = 4/7$, independent of $k$, independent of time, and independent of the initial power spectrum. For a power law initial $P(k)$, $Q_{123}$ is independent of time and of overall scale $k$, but depends on configuration shape in the ratios $k_i/k_j$ and angles $\hat{\boldsymbol{k}}_i \cdot \hat{\boldsymbol{k}}_j$. This dependence of the three-point amplitude on configuration shape is the signature of structure formed by gravitational instability.

### 2.2. Bias

Over the past decade it has become popular to allow for a bias in galaxy formation. Kaiser (1984) introduced the concept of a threshold bias originally to relate the clustering properties of Abell clusters to those of galaxies by examining the statistical properties of those regions where the density contrast exceeds some threshold, $\delta > t$. Expressing the threshold as a multiple of the rms fluctuation, $t = \nu \sigma_0$ (possibly after smoothing the distribution), Kaiser showed that, for $\nu > 1$ and $\xi_0 < 1$, the biased distribution has a correlation function $\xi_b = b_\nu^2 \xi / \xi_0$, with $b_\nu = \nu$. This picture was soon extended to include the possibility that the galaxy distribution is biased relative to the mass density. The implications of a threshold bias have been investigated analytically in great detail for an underlying Gaussian distribution (Kaiser 1984; Politzer & Wise 1984; Bardeen *et al.* 1986; Jensen & Szalay 1986). Jensen & Szalay extended the results to small $\nu$, obtaining the amplification factor

$$b_\nu = \frac{\nu}{F(\nu)}, \qquad F(\nu) = \sqrt{\frac{\pi}{2}} \, \nu \, e^{\nu^2/2} \, \mathrm{erfc}(\nu/\sqrt{2}). \qquad (5)$$

The denominator $F(\nu)$ approaches 1 when $\nu$ is large, thus reducing to the Kaiser result, and vanishes linearly as $\nu \to 0$, such that the factor multiplying $\xi/\xi_0$ is $b_0^2 = 2/\pi$. Many other more or less physically well-motivated models have been considered, with results often summarized by the value of a bias parameter $b$, the ratio of the density contrast in the biased distribution to that in the underlying density, $\delta_b = b\delta$. If a linear bias were the complete story, the spectra would obey $P_b = b^2 P$, $B_b = b^3 B$, and thus $Q_b = Q/b$. However, it seems likely that the process of galaxy formation leads to a galaxy distribution which does not depend linearly on the underlying mass distribution

The threshold model is one example of a nonlinear bias. Applied to an underlying Gaussian, the threshold model gives $Q = 1$ in the weak fluctuation regime (Politzer & Wise 1984; Jensen & Szalay 1986), while for strong clustering $Q \to 1/3$ (Jensen & Szalay 1986). Generally we would expect that the galaxy distribution is a functional of the matter density, $\delta_b(\boldsymbol{x}) = f[\delta(\boldsymbol{x})]$. If the range of effects that determine the efficiency of galaxy formation is not too large, this perhaps can be approximated as an arbitrary but local function, $\delta_b = f(\delta)$. Szalay (1988) has considered an arbitrary biasing function, expanded in Hermite polynomials as $f(\delta) = \sum J_k H_k(\delta)/k!$, applied to a Gaussian underlying density field and obtained a hierarchical three-point function with $Q = J_2/J_1^2$. The next step is to account for a non-Gaussian matter distribution, as that produced by gravitational instability. Fry & Gaztañaga (1993) considered the implications of an arbitrary local bias function applied to an underlying hierarchical mass distribution. Expanding $f(\delta)$ in a series

$$\delta_b = \sum_k \frac{1}{k!} b_k \delta^k,\tag{6}$$

they found, to leading order for small fluctuations, hierarchical galaxy correlations for all orders, with amplitudes that depend on the original $Q_n$ and on the bias parameters $b_n$. In particular, the biased three-point function is hierarchical, with amplitude

$$Q_{123,b} = \frac{1}{b}Q_{123} + \frac{b_2}{b^2}\tag{7}$$

(Fry & Gaztañaga 1993; Juszkiewicz *et al.* 1993), showing a contribution from nonlinear gravity that depends on configuration and a contribution from nonlinear biasing that is independent of configuration. No model of bias that has been suggested to date generates a contribution to $Q$ that depends on configuration, although a more complicated bias model could conceivably do so. Further dependence on configuration shape can arise from higher order moments for non-Gaussian initial conditions (Fry & Scherrer 1994).

## 3. Numerical Results
### 3.1. *The Power Spectrum $P(k)$*

Melott & Shandarin (1993) and Fry, Melott, & Shandarin (1993) have investigated the power spectrum and the bispectrum of the mass distribution for an ensemble of numerical simulations. The results in the perturbative regime agree within uncertainties with gravitational instability predictions in perturbation theory and obey expected scalings in the nonlinear regime up to departures that we believe are understood. In this paper we examine the effects of a threshold bias applied to the same ensemble of models. The simulations are performed on a staggered $128^3$ mesh (Melott 1986), starting from small amplitude Gaussian initial conditions with random phases for the Fourier components. For the differencing scheme used, growth rates of fluctuations are suppressed for wavenumbers larger than $k > k_{\rm Ny}/2 = 32$ (considerably better than the usual non-staggered mesh value $\sim k_{\rm Ny}/4$). We present results for wavenumbers in the range $1 \leq k \leq 32$, where $k = 1$ corresponds to the fundamental mode.

Models are identified by initial power spectrum index, $P_0 \sim k^n$, for $n = +1$, $0$, $-1$, and $-2$ out to the Nyquist frequency $k_{\rm Ny} = 64$. We have also performed simulations for a spectrum with $n = -3$, but as our earlier results bear out, results of any attempt to realize such a divergent spectrum on a finite volume are severely affected by the cutoff, and so in this paper we examine only $n \geq -2$. For each initial spectrum we have information for a succession of evolution stages labeled by $k_{\rm nl}$, the scale of nonlinearity given by

$$\int_0^{k_{\rm nl}} d^3k\, P(k) = 1,\tag{8}$$

evaluated for the analytically evolved initial spectrum, $P = a^2(t)P_0(k)$. A given $k_{nl}$ represents evolution for an expansion factor of approximately $a = \Delta_0^{-1}(k_{Ny}/k_{nl})^{(3+n)/2}$, where $\Delta_0$ is the initial rms fluctuation in one cell.

For each initial spectrum and for each stage of evolution we construct a "biased" distribution by keeping only particles in cells with density above an absolute threshold $\delta > t$ on the scale of 1 cell at three different threshold levels, $t = 3$, $t = 10$, and $t = 30$. Table 1 provides the rms cell fluctuation $\sigma_0$ at each stage necessary to translate this threshold into a number of standard deviations $t = \nu\sigma_0$. At evolution stage $k_{nl} = 32$ fluctuations on the scale of 1 cell are just becoming nonlinear ($\sigma_0 = 1.2 - 1.8$), while at stage $k_{nl} = 4$ the degree of nonlinearity is large ($\sigma_0 = 8.4 - 12.3$).

After applying the threshold, unlike the situation in our earlier analyses of the unbiased distribution, the number of particles remaining is so small that we now must remove discreteness contributions to the the Fourier spectra,

$$\langle |\tilde{\delta}(\boldsymbol{k})|^2 \rangle = P(k) + \frac{1}{N}, \tag{9}$$

$$\langle \tilde{\delta}_1 \tilde{\delta}_2 \tilde{\delta}_3 \rangle = B_{123} + \frac{1}{N}[P(k_1) + P(k_2) + P(k_3)] + \frac{1}{N^2}, \tag{10}$$

where $P$ and $B$ are the contributions from clustering. The numerical data suggest the appropriate discreteness measure is the number $N = N_c$ of cells occupied above threshold. For comparison with later results, Table 2 shows the discreteness correction $P_d = 1/N_c$ averaged over two random realizations for $k_{nl} = 32$ and $k_{nl} = 4$ (the earliest and latest evolution stages), as a function of spectra index $n$ and threshold $t$. For late stages of evolution at lower thresholds, $N$ is large, and the discreteness correction is insignificant. At early stages of evolution and higher thresholds, the correction can be large. For the most extreme case, $n = +1$, $k_{nl} = 32$, $t = 30$, we have for two random realizations $N_c = 3$ and $N_c = 0$. It is likely in this case that the excess power above discreteness is just random noise; we include it because there may be information in the distribution of this noise.

Figures 1–4 show $P(k)$ after the correction for discreteness. Error flags in the figures show the dispersion between random realizations, except for $k_{nl} = 32$, $t = 30$, where the error indicates the variation within the one realization that has $N_c \neq 0$. Recall that early stages with large thresholds are dominated by discreteness and are not likely to be reliable. Figures 1 and 2 both show results for initial spectrum $n = -1$. Within each window in Figure 1 the four curves show the effect of increasing threshold at the same evolution stage, as indicated. At an early evolution stage ($k_{nl} = 32$) the power depends strongly on the threshold; however his dependence steadily diminishes with evolution, becoming almost negligible at $k_{nl} = 4$. Qualitatively, this behavior could be anticipated: at later stages a considerable amount of mass is in dense clumps, and increasing the threshold makes almost no difference in the mass distribution. The four windows in Figure 2 show increasing threshold, from no bias to $t = 30$, while the curves within each window show stages of evolution. The effect mentioned above is demonstrated in a different form in Figure 2, where the effect of time evolution steadily diminishes with increasing threshold. That the short wave part of the spectrum of the highest peaks ($\delta > 30$) appears to decrease with time is probably an artifact of the shot noise subtraction, which dominates the power here. Figures 3 and 4 show the effect of increasing threshold for the four values of initial spectral index, for $k_{nl} = 32$ (Fig. 3) and at $k_{nl} = 4$ (Fig. 4). The dramatic difference between Figure 3 and Figure 4 illustrates the influence of the threshold bias in the mildly nonlinear regime ($\sigma_0 = 1.2$–$1.8$ in Fig. 3) and highly nonlinear regime ($\sigma_0 = 8.4$–$12.3$ in Fig. 4).

These results follow the qualitative trend that a larger bias threshold produces a greater amplification of the power spectrum. Surprisingly, the linear bias prediction $P_b(k) = b^2 P(k)$, is at least approximately quantitatively valid as well. The first column under each heading in Table 3 shows the value of $b$ derived from the power spectrum averaged for $1 \leq k \leq 32$ (this is not necessarily a good fit) and the second the Jensen & Szalay (1986) predicted amplification factor $b_\nu$ for the given threshold (eq. [5]). Comparing the two columns one can see a general correspondence, although a difference of a factor of 2 or greater is not unusual. Note that this agreement obtains only if we omit the factor $\sigma_0$ that appears in the denominator of the perturbation theory prediction.

## 3.2. The Three-Point Amplitude $Q_{123}$

We next present the reduced three-point amplitude $Q_{123}$. Figures 5–8 show $Q(k)$ for equilateral triangle configurations after the discreteness correction in a manner similar to the power spectrum. For clarity, errors are shown only for selected curves; in general, errors are comparable to the apparent fluctuation from point to point. As described in § 2, the gravitational instability prediction for the quasi-linear regime is $Q = 4/7$, indicated as the horizontal dashed lines on the left of each panel; and numerical simulations suggest $Q(n) \approx 3/(3 + n)$ for no bias in the nonlinear regime, indicated by the horizontal dashed lines on the right. As in Figure 1, the four windows in Figure 5 show increasing evolution for $n = -1$, from $k_{\rm nl} = 32$ to $k_{\rm nl} = 4$. The curves in each window show increasing threshold, as labeled in the caption. In the quasi-linear regime $k < k_{\rm nl}$, $Q \approx 4/7$ appears to hold within the errors (which are large for small $k$). ¿From equation (7), this is expected for no bias but also holds for certain combinations of bias parameters; for instance, for $b \gg 1$ if $b_2/b^2 \approx 4/7$. For $k > k_{\rm nl}$, $Q$ steadily grows, towards the expected $Q = 3/2$ with no bias, but with a mild but consistent decrease with increasing bias, closer to $Q = 1$ for $t = 30$. The four windows in Figure 6 show $n = -1$ with increasing threshold, from no bias to $\delta > 30$; the curves in each window show increasing evolution, as labeled in the caption. With no bias the rise from $Q = 4/7$ towards $Q = 3/2$ is noticeable, especially at late evolution, while with the largest bias $t = 30$, $Q = 1$ for scales in the nonlinear regime.

The windows in Figures 7 and 8 show increasing bias for different values of initial spectral index, from $n = +1$ to $n = -2$, at $k_{\rm nl} = 32$ (Fig. 7) and at $k_{\rm nl} = 4$ (Fig. 8). Fry, Melott, & Shandarin (1993) show that errors in $Q$ go as $(k_{\rm nl}/k)^{(3+n)}$, and thus are very large for positive $n$ and small $k$; points with $\Delta Q \gg Q$ are not shown. Figure 7 shows an early stage when all scales shown are in the quasi-linear or mildly nonlinear regime. Therefore we do not see much departure from $Q = 4/7$, except for $n = -2$ with no bias. In Figure 8, on the other hand, all scales except the smallest $k$ are in the strongly nonlinear regime, and we see a steady growth of $Q$ from $Q = 4/7$ for small $k$ towards $Q(n) = 3/(3 + n)$ at large $k$ without bias or $Q \approx 1$ at high threshold. The models dominated by the small scale perturbations, $n = +1$ and $n = 0$, do not display much dependence on the bias factor; however, in the $n = -2$ model this dependence is quite significant. Figure 9 shows one final view of different parameter combinations, four different values of the initial spectral index $n$ in windows of increasing bias threshold. As shown in Figure 9, with no bias $Q$ approaches the corresponding value $Q(n) = n/(n + 3)$ at $k = 32$; with the largest bias, $\delta > 30$, all curves look similar and approach the value $Q = 1$ at $k = 32$. The no bias window also shows $k_{\rm nl} = 8$.

Finally, we examine the dependence of $Q_{123}$ on configuration shape. We plot in Figure 10 the reduced three-point amplitude $Q(\theta)$ for configurations with $k_1 = 32$ and $k_2 = 16$, separated by angle $\theta$, for initial $n = -1$. For clarity error flags are shown only for no bias and for $t = 3$. The earliest stage shows a substantial variation of $Q$ with $\theta$, with enhanced correlation for $\theta = 0$ and $\theta = \pi$. One can intuitively understand the origin of such a shape dependence from gravitational instability. The Zel'dovich (1970) approximation describes pancaking (highly anisotropic collapse) from initial conditions with negligible power at large $k$ in the initial conditions. Coles, Melott, & Shandarin (1993; see also Melott, Pellman, &

Shandarin 1994) showed that a simple extension of the Zel'dovich approximation applies to Gaussian initial conditions with a wide variety of power spectra. Thus the enhancement of local anisotropy is a generic result of gravitational instability. In transform space, the effect of a two-dimensional spatial density enhancement is a preference for wave vectors orthogonal to the two-dimensional structure, and in statistics averaged over orientations this survives as an enhancement for nearly colinear configurations. The clumpiness of pancakes and filaments eliminates this effect at late stages. Since the size of the triangle in the transform space ($k_1 = 32$ and $k_2 = 16$) remains constant during the evolution, we probe scales a little greater than the size of clumps at the stages $k_{nl} = 32$ and 16 and see enhancement of $Q$ at small $\theta$ for $n = -2$. On the other hand at later stages $k_{nl} = 4$ the triangle probes the interior of the clumps and the enhancement disappears.

## 4. Discussion

In this paper we have examined the power spectrum and bispectrum for an ensemble of numerical simulations with an applied threshold bias. As predicted in perturbation theory we have seen that that the general effect of a threshold bias on the power spectrum is to amplify the power by a roughly constant factor that can be written as $b^2$; this result applies from the quasi-linear regime into the strongly nonlinear regime. The numerical results follow qualitatively the trend predicted in the amplification factor $b_\nu$ of equation (5) (Jensen & Szalay 1986), but depart from exact numerical agreement by factors of 2 or larger. A more serious departure from the prediction is that agreement is achieved only when the factor $\xi_0$ in the prediction $\xi_g = b_\nu^2 \xi / \xi_0$ is set to 1. Yet it is perhaps the agreement rather than the departure that is the more surprising, because many of the numerical results are far outside the domain where perturbation theory should apply.

The effect of bias on the three-point function is different in the quasi-linear and strongly nonlinear regimes. In the nonlinear regime the effect of bias on the amplitude $Q(k)$ for equilateral triangles is to modify the value $Q(n) = 3/(3+n)$ for no bias to $Q \approx 1$, independent of $n$, for high bias at large $k$. This confirms the result that biased $Q \approx 1$ found previously by Melott & Fry (1986) for simulations on a much coarser grid. This value is expected for the threshold bias when $b \gg 1$, since effectively $b_2/b^2 = 1$, but it appears to hold for large threshold even when the amplification $b$ is not much different from 1. In the quasi-linear regime, where we ought to be able to make reliable predictions, the errors become large and it is difficult to draw specific conclusions. It seems that $Q(k)$ again approaches a constant value for small $k$, but it would be difficult to conclude that this value is different from the gravitational instability result $Q = 4/7$. Note that even so, for this to be consistent with equation (7) with a large $b$ inferred from the power spectrum $P(k)$, a $b_2$ term is necessary.

Further information can be extracted from the dependence of the three-point amplitude $Q$ on configuration shape. Results for $k_{nl} = 32$ are the least evolved and thus should be the closest to perturbation theory predictions. The perturbation theory prediction depends on spectral index, which evolves slightly from the initial value $n = -1$ and depends slightly on scale: for small wavenumbers, $1 \leq k \leq 16$, $n = -1.2$; while for $16 \leq k \leq 32$, $n = -1.6$. For $t = 3$, fitting $Q(\theta)$ to the form in equation (7) with $n = -1$ gives $b = 4.5 \pm 0.2$, $b_2/b^2 = 0.51 \pm 0.01$. For $n = -1.5$, a value we might infer from $P(k)$ at the stage observed, $b = 5.5 \pm 0.3$, $b_2/b^2 = 0.52 \pm 0.01$. For $t > 10$, a fit with $n = -1$ gives $b = 6.9 \pm 0.6$, $b_2/b^2 = 0.67 \pm 0.01$; $n = -1.5$ gives $b = 8.1 \pm 0.8$ and $b_2/b^2 = 0.68 \pm 0.01$. (These errors are formal errors in a linear least squares fit, ignoring parameter correlations.) For comparison, the ratio of the biased and unbiased power spectrum in these two cases over the same range in $k$ gives $b = 4.10 \pm 0.01$ and $b = 9.5 \pm 0.2$. It appears that the shape dependence of the three-point function slightly overestimates the amount of bias, but we note that even for $k_{\rm nl}$ 32, the results have evolved beyond the truly perturbative regime. In the highly evolved results, $k_{\rm nl} = 4$, there is less dependence of $Q$ on configuration shape, even with no bias. What dependence $Q$ has on $\theta$ is very different from that in perturbation theory, but it appears to be robust and to be affected by bias only in amplitude, with no offset $[b_2 = 0$ in eq. (7)]. Using the form $Q_b = Q/b$ applied to the curves for $k_{\rm nl} = 4$ in Figure 10 gives $b = 1.24$ for $t = 3$, $b = 1.36$ for $t = 10$, and $b = 1.54$ for $t = 30$, values that compare well with those obtained from the power spectrum in Table 3.

In conclusion, we have seen that a threshold bias, even at a very high level, changes the power spectrum and bispectrum of the biased distribution, but in ways that make it still possible to extract useful information about the underlying matter distribution and about initial conditions. Bias does not change the shape of the power spectrum except when the threshold is very large. In the nonlinear regime, with almost any appreciable bias $Q \approx 1$ for equilateral triangles, independent of initial spectrum or degree of evolution. However, in the perturbative regime, $k < k_{\rm nl}$, the characteristic dependence of the bispectrum on the shape of the triangular configuration in perturbation theory (eqs. [2–3]) allows in principle the separation of the contributions of bias and gravity, and provides an important new, totally independent method for estimating $b$ that does not require separate knowledge of the amplitude of the unbiased power spectrum.

Research supported in part by NASA grant NAGW-2381 and by the DOE at Fermilab, and by NASA grant NAGW-3832 and National Science Foundation grants AST-9021414 and OST-9255223 at the University of Kansas. ALM is grateful to the Institute of Astronomy, Cambridge University, and JNF to Fermilab for their hospitality during the period in which some of this work was done. We thank the National Center for Supercomputing Applications, Urbana, Illinois for the grant of Convex C3 time which made this work possible.

Table 1
R.M.S. Cell Fluctuation $\sigma_0$

| $n$ | $k_{nl} = 32$ | $k_{nl} = 16$ | $k_{nl} = 8$ | $k_{nl} = 4$ |
|---|---|---|---|---|
| +1 ...... | 1.20 | 3.06 | 6.42 | 12.30 |
| 0 ...... | 1.22 | 2.95 | 5.91 | 10.74 |
| −1 ...... | 1.22 | 2.67 | 5.83 | 9.83 |
| −2 ...... | 1.84 | 3.19 | 5.16 | 8.36 |

Table 2
Discreteness Power $P_d = 1/N_c$

| $k_{nl}$ | $t$ | $n = +1$ | $n = 0$ | $n = -1$ | $n = -2$ |
|---|---|---|---|---|---|
| 32 ........ | 3 | $1.65 \times 10^{-5}$ | $1.33 \times 10^{-5}$ | $1.92 \times 10^{-5}$ | $1.50 \times 10^{-5}$ |
| | 10 | $5.02 \times 10^{-4}$ | $2.20 \times 10^{-4}$ | $2.95 \times 10^{-4}$ | $1.14 \times 10^{-4}$ |
| | 30 | $3.33 \times 10^{-1}$ | $4.63 \times 10^{-2}$ | $1.25 \times 10^{-2}$ | $1.31 \times 10^{-3}$ |
| 4 ........ | 3 | $1.86 \times 10^{-5}$ | $1.60 \times 10^{-5}$ | $1.52 \times 10^{-5}$ | $1.26 \times 10^{-5}$ |
| | 10 | $3.26 \times 10^{-5}$ | $3.03 \times 10^{-5}$ | $3.32 \times 10^{-5}$ | $3.80 \times 10^{-5}$ |
| | 30 | $6.98 \times 10^{-5}$ | $7.15 \times 10^{-5}$ | $8.48 \times 10^{-5}$ | $1.19 \times 10^{-4}$ |

Table 3
Bias Amplification Factor $b$ ($n = -1$)

| $t$ | $k_{nl} = 32$ | | $k_{nl} = 16$ | | $k_{nl} = 8$ | | $k_{nl} = 4$ | |
|---|---|---|---|---|---|---|---|---|
| 3 ...... | 4.1[a] | 2.9[b] | 2.1 | 1.6 | 1.4 | 1.2 | 1.2 | 1.0 |
| 10 ...... | 9.1 | 8.3 | 3.2 | 4.0 | 1.8 | 2.1 | 1.5 | 1.5 |
| 30 ...... | 24.1 | 24.6 | 5.2 | 11.3 | 2.5 | 5.8 | 1.7 | 3.3 |

[a] Observed $b$ from $P(k)$

[b] Predicted $b_\nu$ from equation (5)

Figures

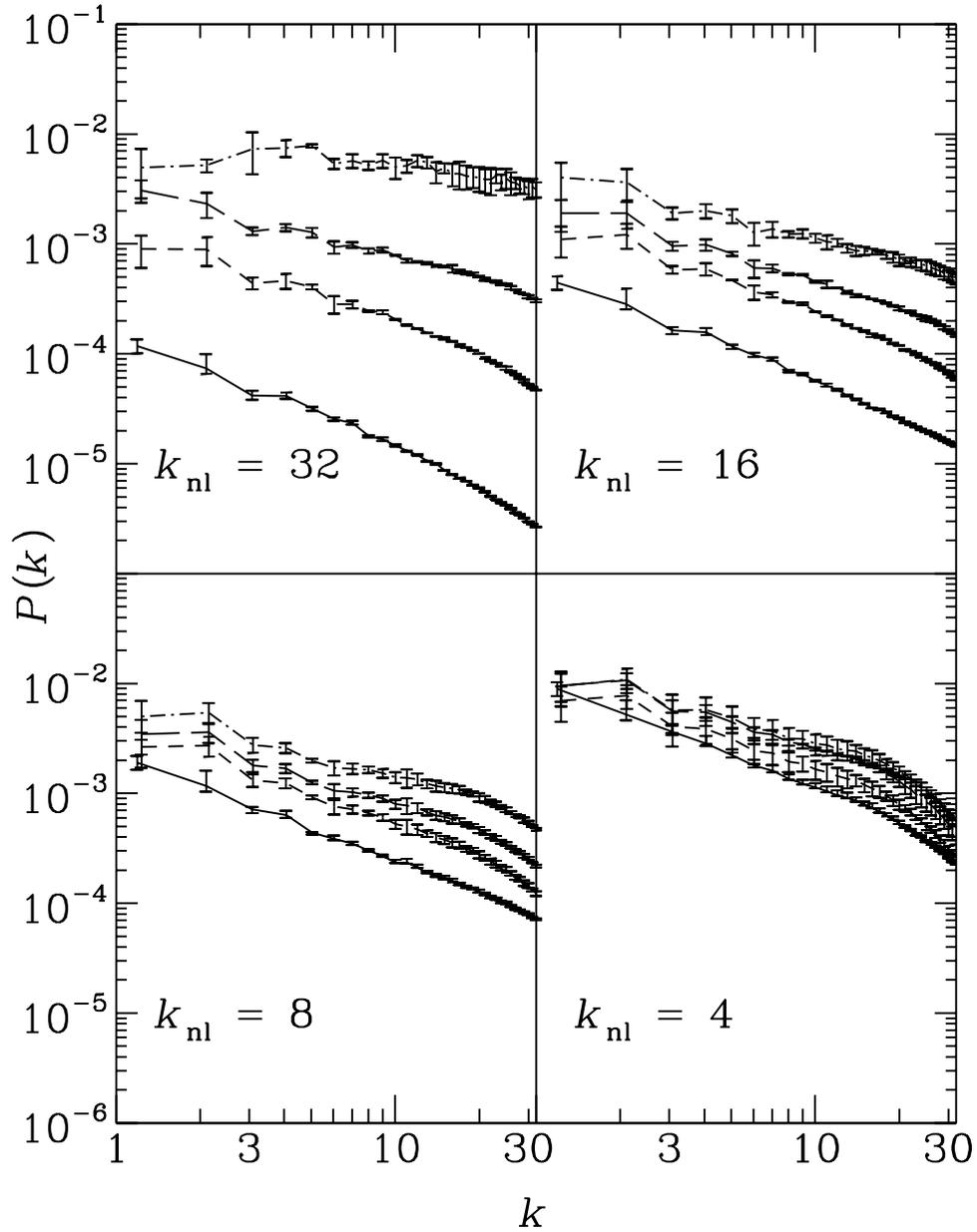

Fig. 1.—$P(k)$ vs. $k$ for models with $n = -1$. The four windows show evolution stages $k_{nl} = 32$, $k_{nl} = 16$, $k_{nl} = 8$, and $k_{nl} = 4$. In each window curves are plotted for no bias (solid line), and for bias thresholds $\delta > 3$ (short-dash line), $\delta > 10$ (long-dash line), and $\delta > 30$ (dot-dash line).

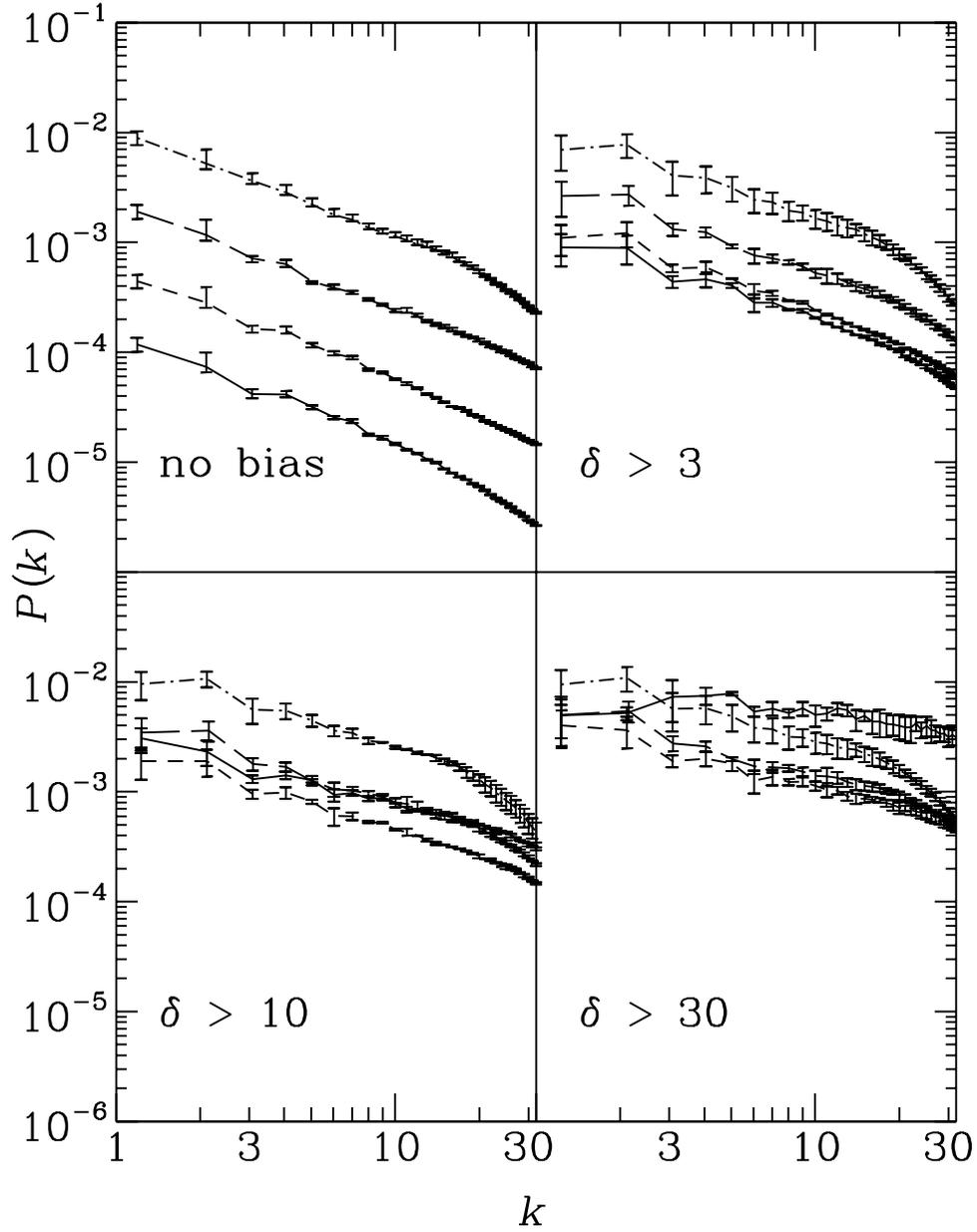

Fig. 2.—$P(k)$ vs. $k$ for models with $n = -1$. The four windows show no bias and bias thresholds $\delta > 3$, $\delta > 10$, and $\delta > 30$. In each window curves are plotted for evolution stages $k_{\mathrm{nl}} = 32$ (solid line), $k_{\mathrm{nl}} = 16$ (short-dash line), $k_{\mathrm{nl}} = 8$ (long-dash line), and $k_{\mathrm{nl}} = 4$ (dot-dash line).

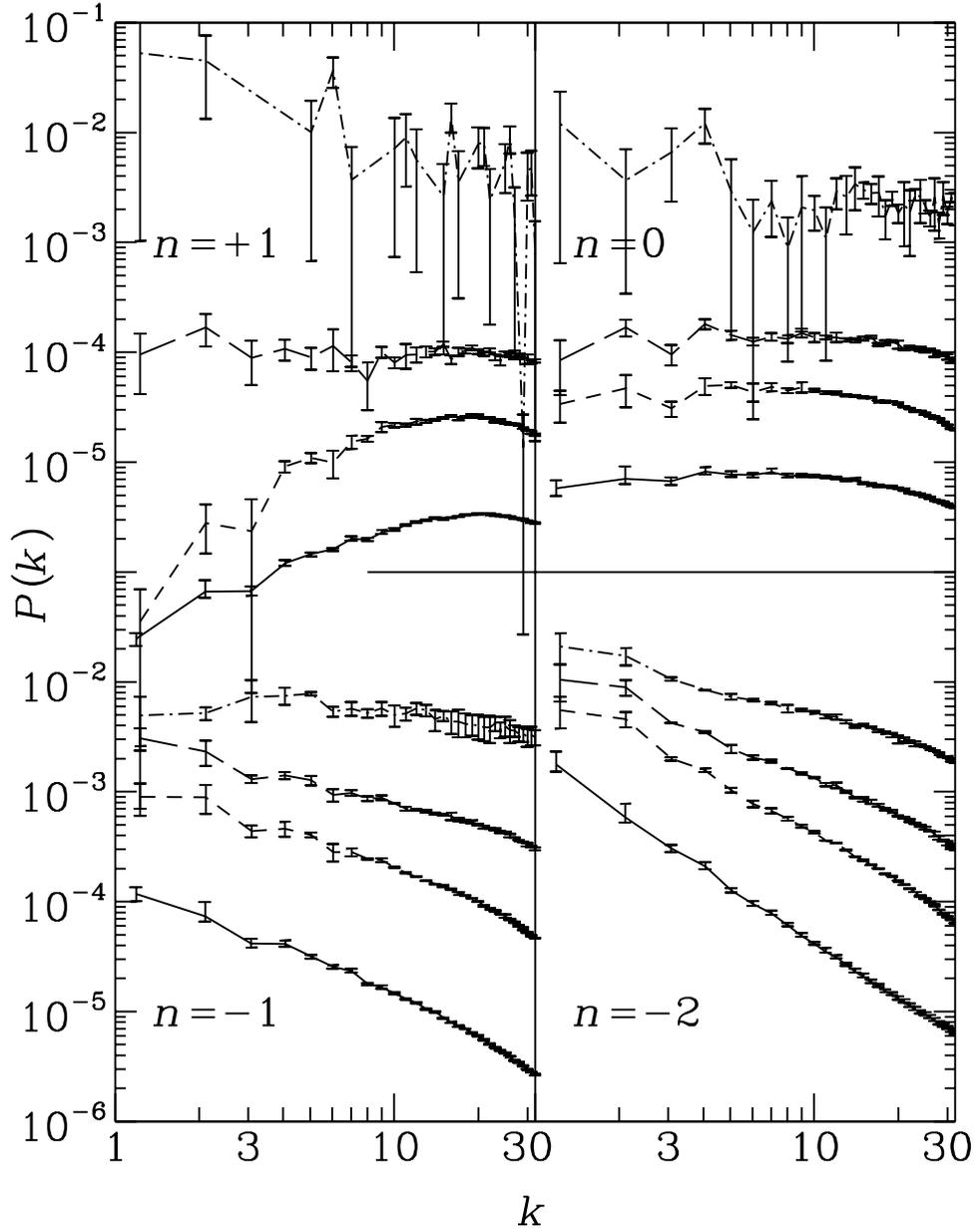

Fig. 3.—$P(k)$ vs. $k$ for models at evolution stage $k_{nl} = 32$. The four windows show initial power spectrum index $n = +1$, $n = 0$, $n = -1$, and $n = -2$. In each window curves are plotted for no bias and for bias thresholds $\delta > 3$, $\delta > 10$, and $\delta > 30$, indicated as in Figure 1.

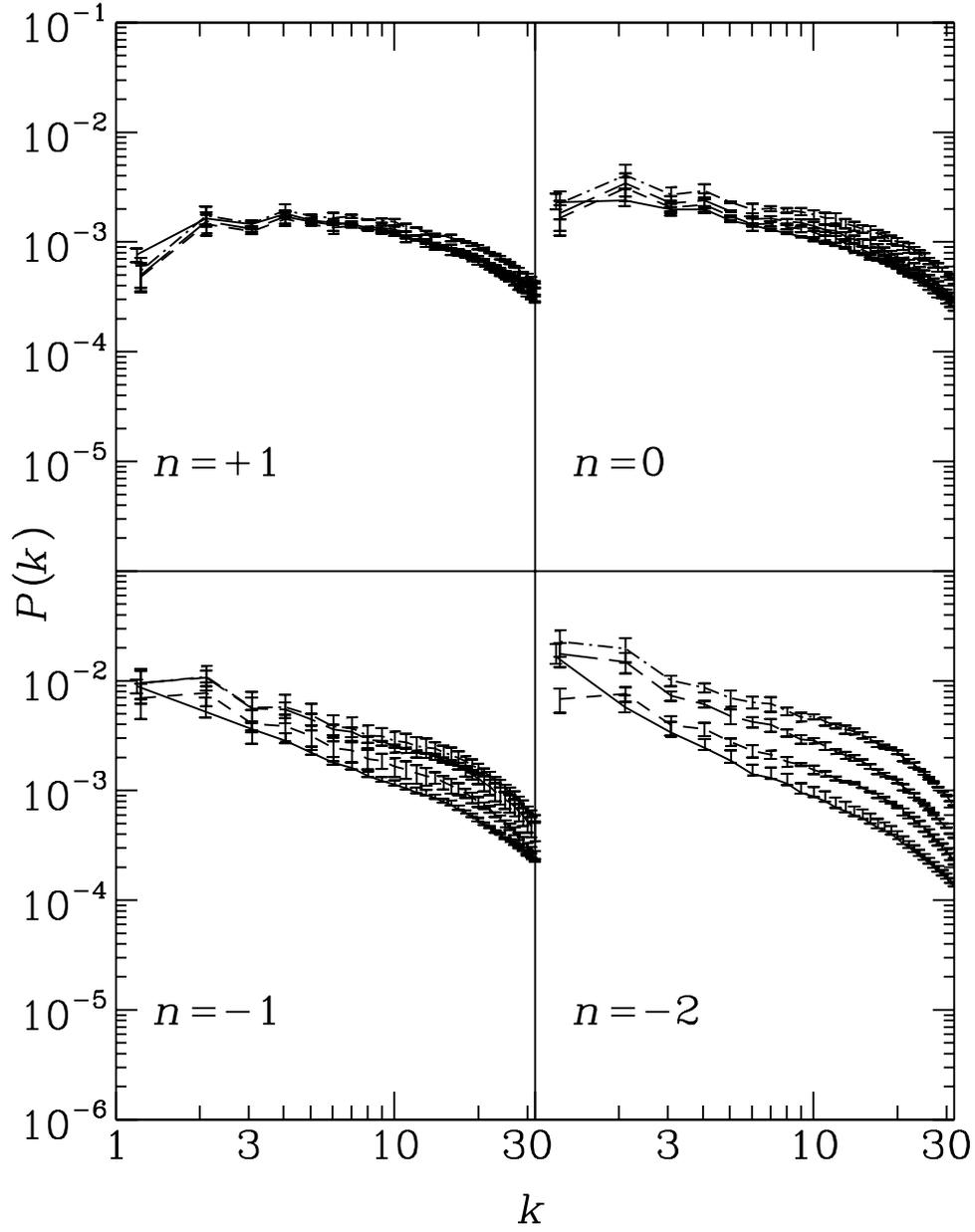

Fig. 4.—$P(k)$ vs. $k$ for models at evolution stage $k_{\rm nl} = 4$. The four windows show initial power spectrum index $n = +1$, $n = 0$, $n = -1$, and $n = -2$. In each window curves are plotted for no bias and for bias thresholds $\delta > 3$, $\delta > 10$, and $\delta > 30$, as in Figure 1.

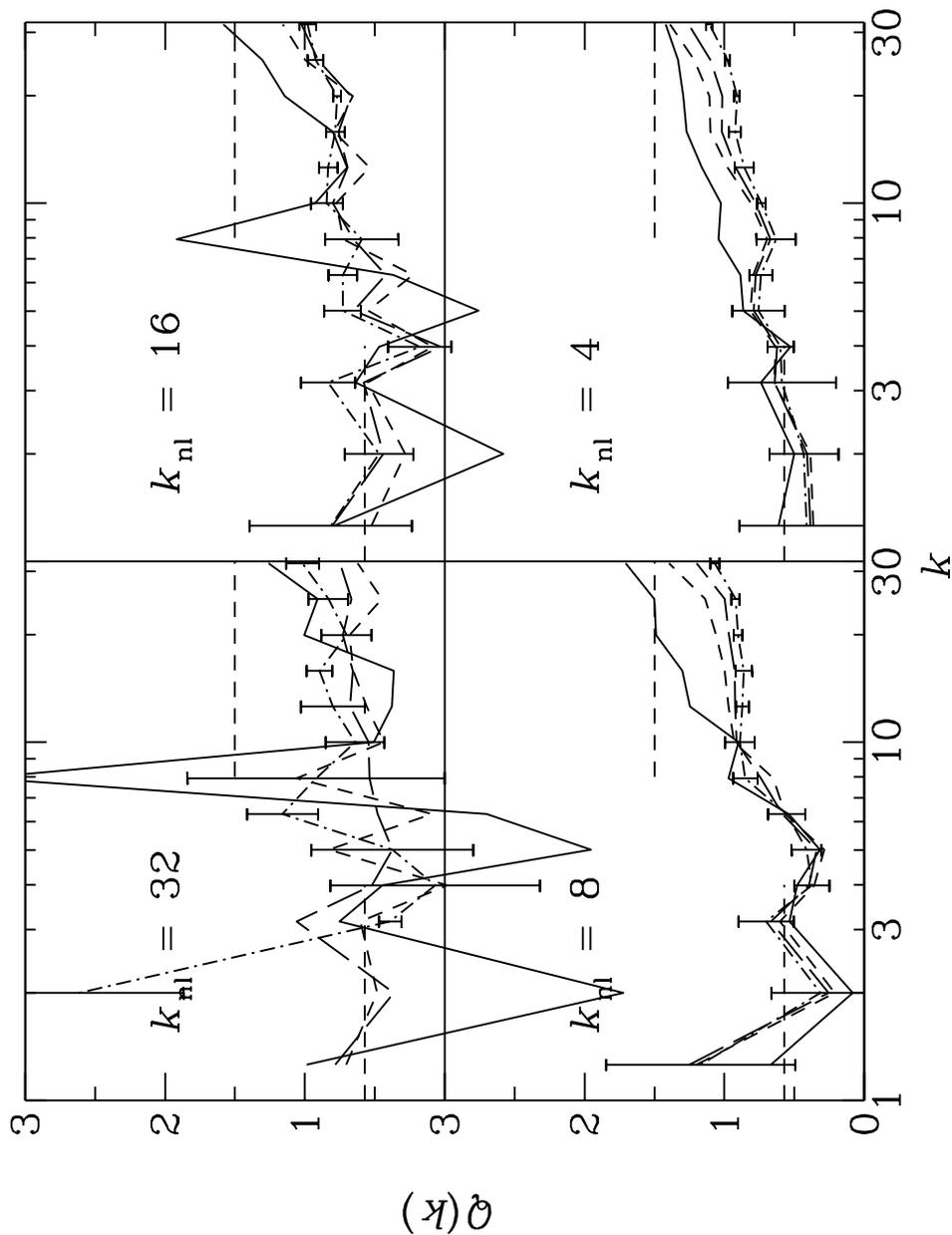

Fig. 5.—$Q(k)$ vs. $k$ for models with $n = -1$. The four windows show evolution stages $k_{nl} = 32$, $k_{nl} = 16$, $k_{nl} = 8$, and $k_{nl} = 4$. In each window curves are plotted for no bias (solid) and for bias thresholds $\delta > 3$ (short-dash), $\delta > 10$ (long-dash), and $\delta > 30$ (dot-dash). Horizontal dashed lines show $Q = 4/7$ expected from perturbation theory and $Q = 3/2$ expected in the nonlinear regime with no bias. For clarity error bars are shown only on one curve in each panel.

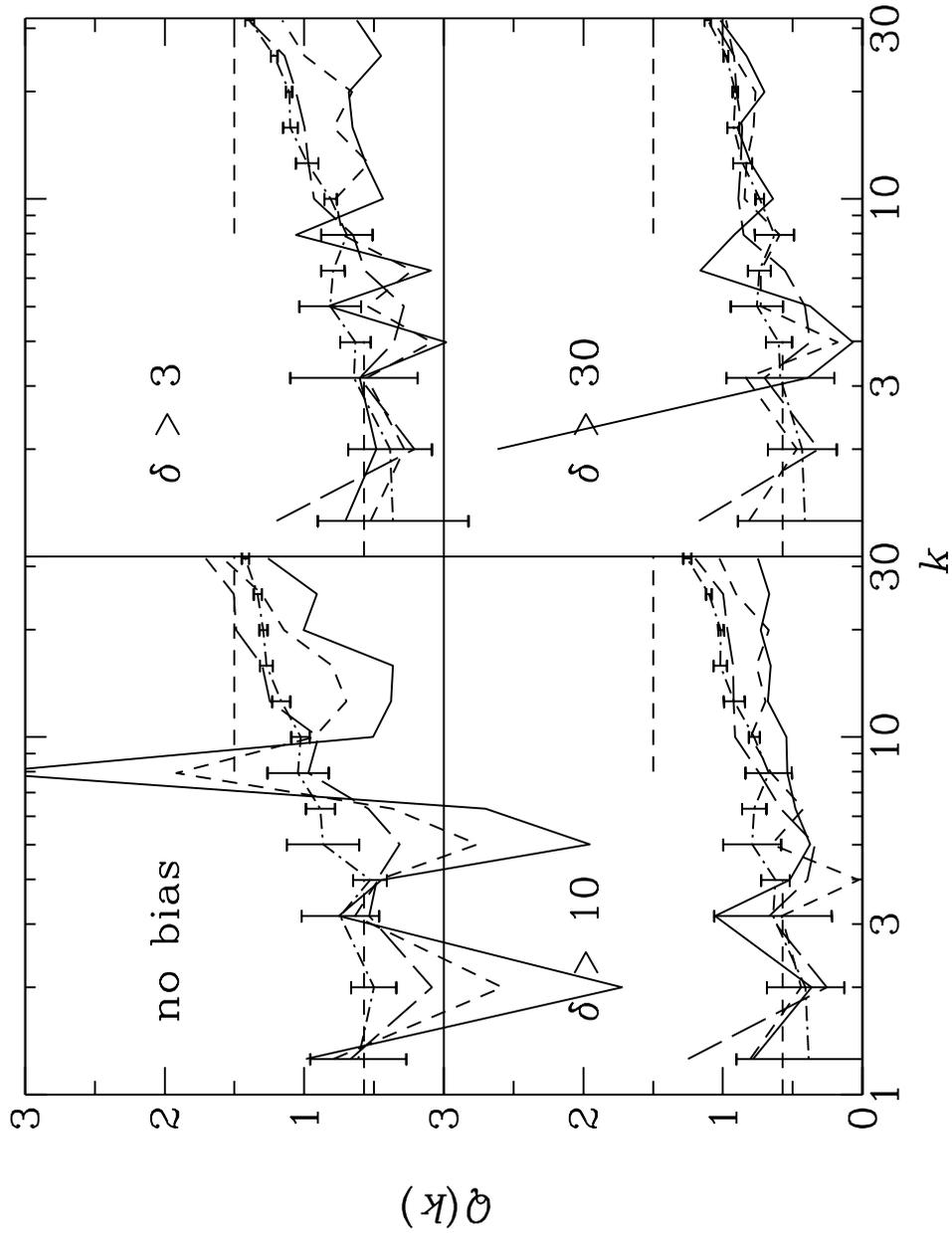

Fig. 6.—$Q(k)$ vs. $k$ for models with $n = -1$. The four windows show no bias and bias thresholds $\delta > 3$, $\delta > 10$, and $\delta > 30$. In each window curves are plotted for evolution stages $k_{nl} = 32$ (solid), $k_{nl} = 16$ (short-dash), $k_{nl} = 8$ (long-dash), and $k_{nl} = 4$ (dot-dash).

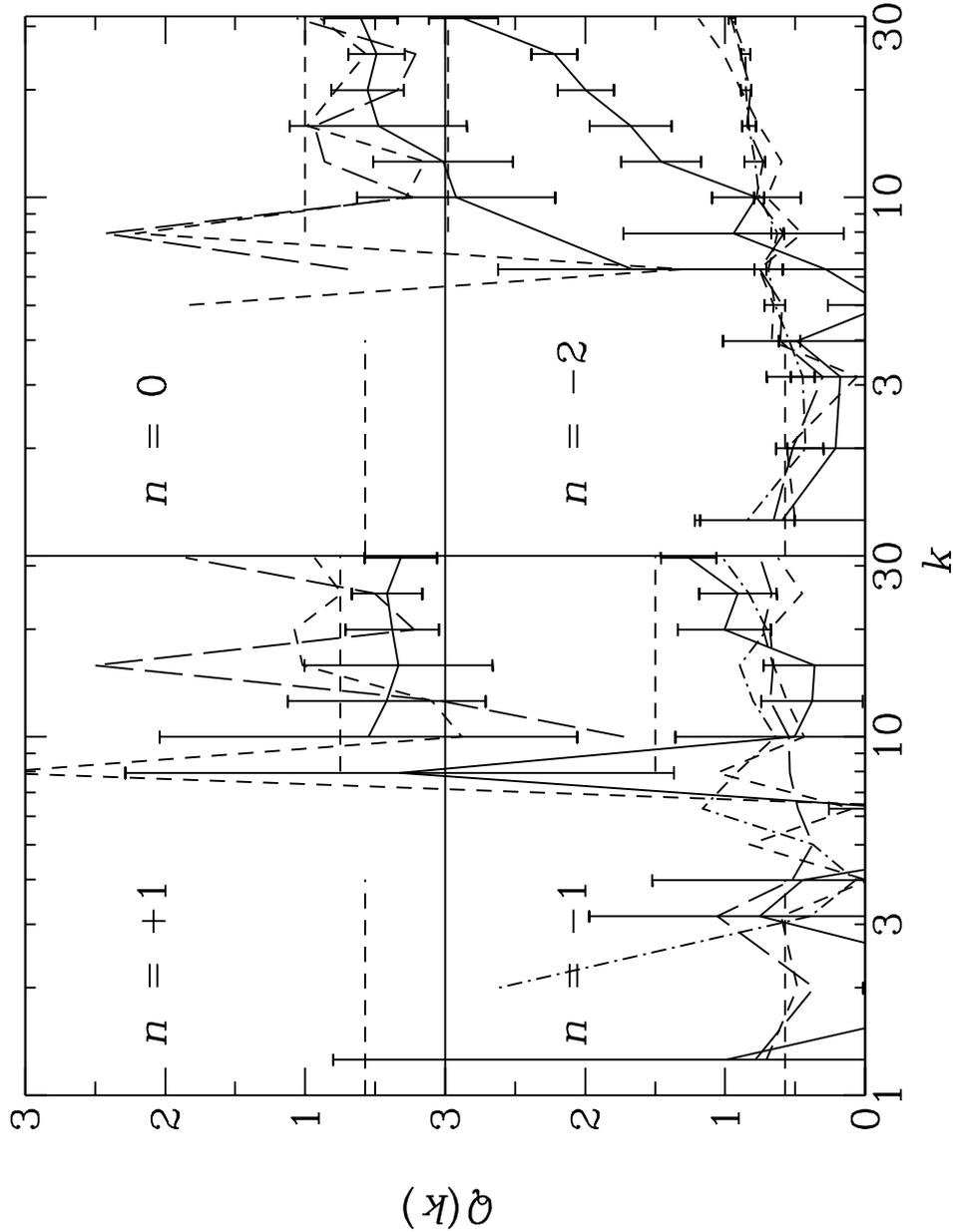

Fig. 7.—$Q(k)$ vs. $k$ for models at evolution stage $k_{\mathrm{nl}} = 32$. The four windows show initial power spectrum index $n = +1$, $n = 0$, $n = -1$, and $n = -2$. In each window curves are plotted for no bias and for bias thresholds $\delta > 3$, $\delta > 10$, and $\delta > 30$ as indicated in Figure 5. Horizontal dashed lines on the right-hand side of each panel show $Q = 3/(n+3)$ expected in the nonlinear regime with no bias.

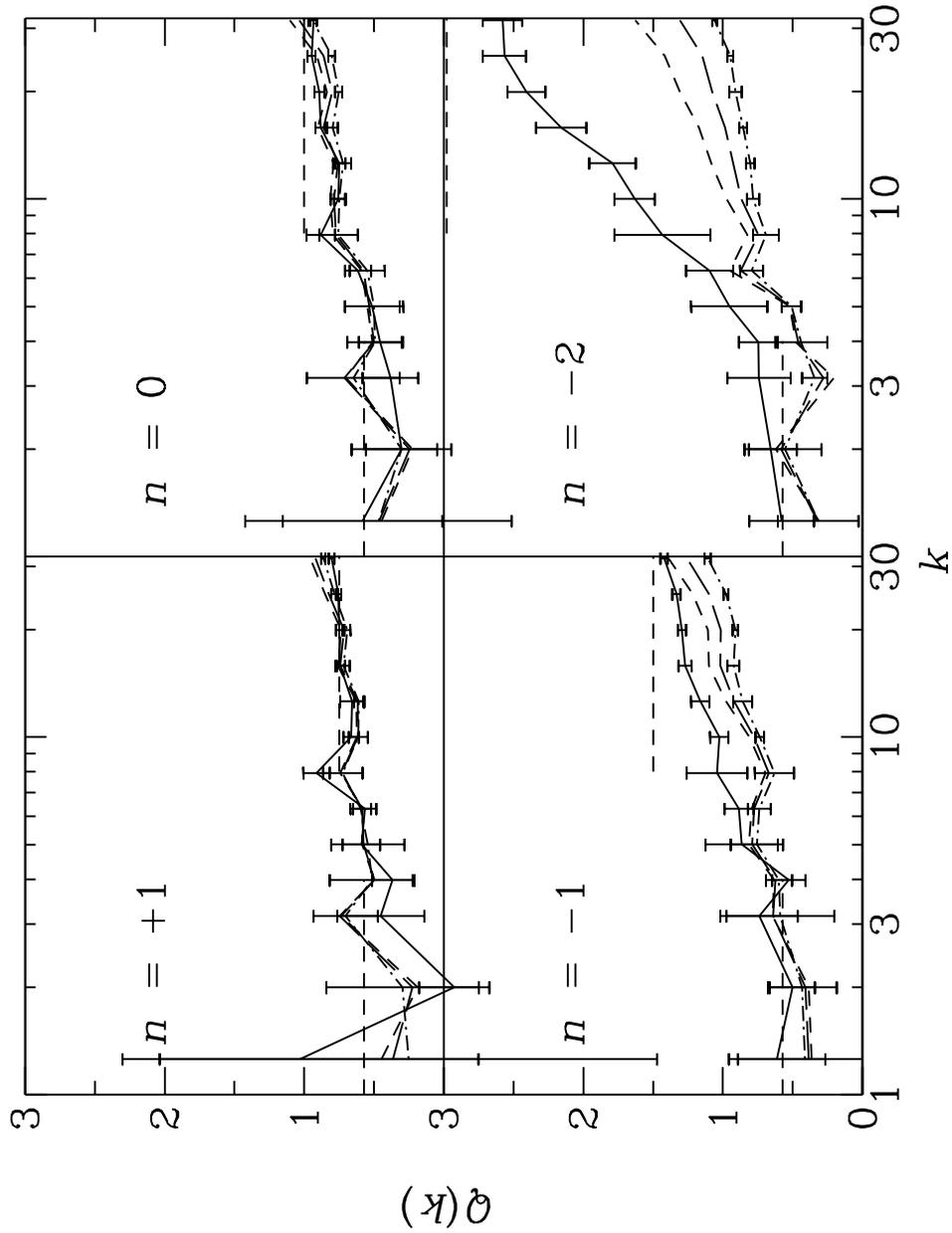

Fig. 8.—$Q(k)$ vs. $k$ for models at evolution stage $k_{nl} = 4$. The four windows show initial power spectrum index $n = +1$, $n = 0$, $n = -1$, and $n = -2$. In each window curves are plotted for no bias and for bias thresholds $\delta > 3$, $\delta > 10$, and $\delta > 30$, as indicated in Figure 5.

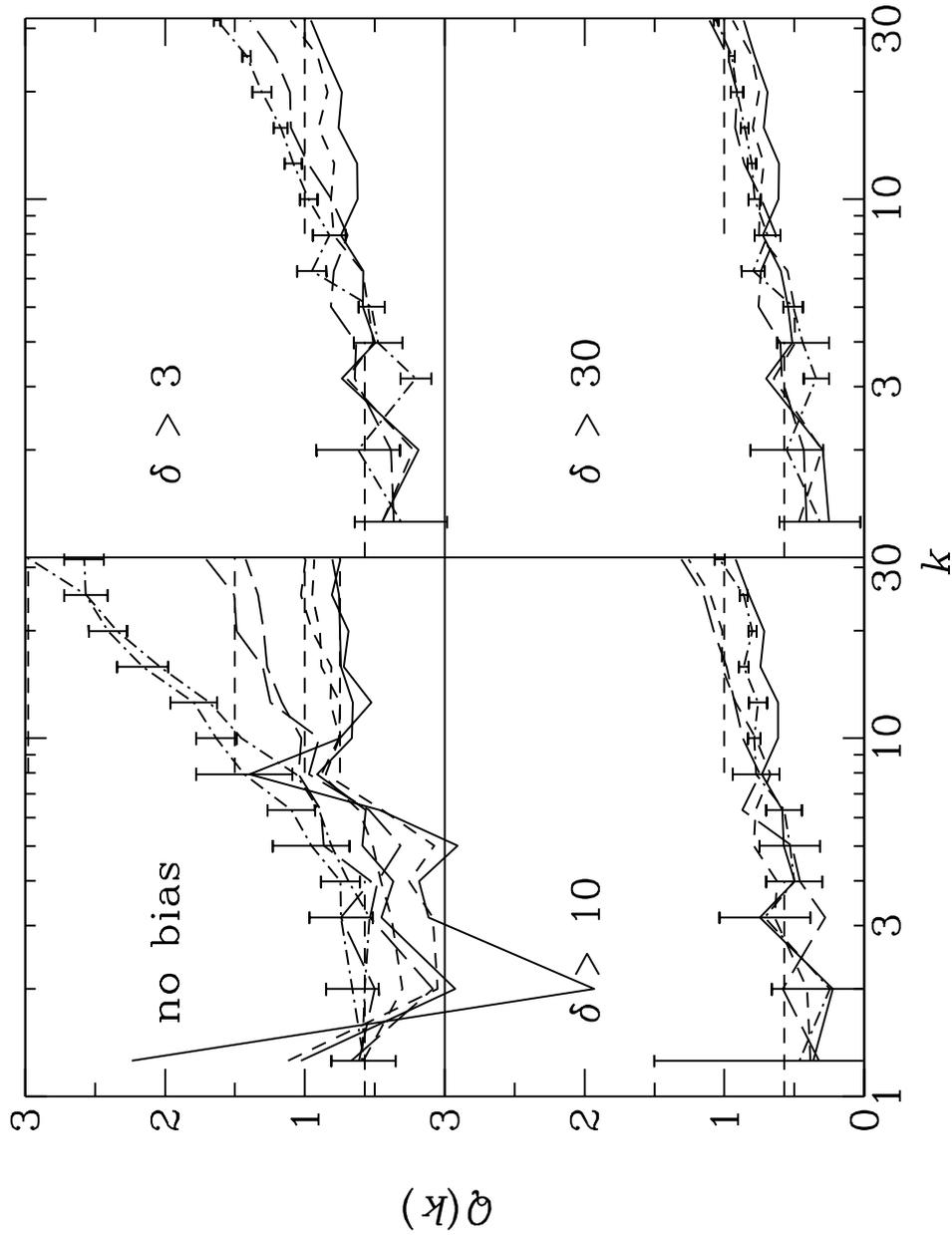

Fig. 9.—$Q(k)$ vs. $k$ for models at evolution stage $k_{\rm nl} = 4$. The four windows show no bias and bias thresholds $\delta > 3$, $\delta > 10$, and $\delta > 30$. In each window curves are plotted for $n = +1$ (solid line), $n = 0$ (short-dash line), $n = -1$ (long-dash line), and $n = -2$ (dot-dash line). The no bias window also shows $k_{\rm nl} = 8$. Horizontal dashed lines on the right side of the no bias panel show the expected asymptotic value $Q(n) = 3/(3+n)$. In the remaining panels $Q = 1$ is marked.

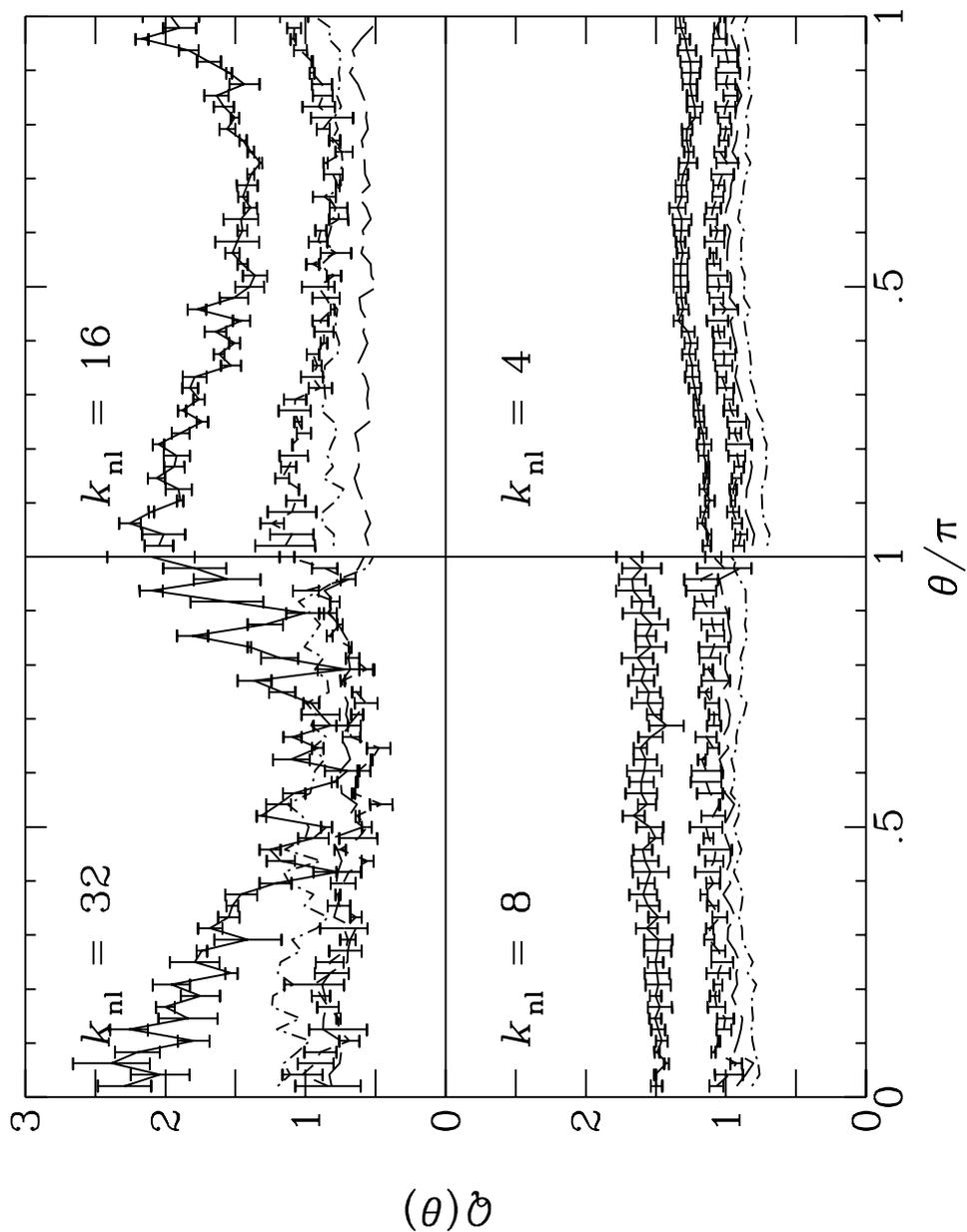

Fig. 10.—$Q(\theta)$ vs. $\theta$ for models with $n = -1$, configurations with $k_1 = 32$, $k_2 = 16$, separated by angle $\theta$ such that $k_3^2 = k_1^2 + k_2^2 + 2k_1k_2\cos\theta$. The four windows show evolution stages $k_{\rm nl} = 32$, $k_{\rm nl} = 16$, $k_{\rm nl} = 8$, and $k_{\rm nl} = 4$. In each window curves are plotted for no bias (solid line) and for bias thresholds $\delta > 3$ (short-dash line), $\delta > 10$ (long-dash line), and $\delta > 30$ (dot-dash line).